# Impacts of the half-skyrmion spin topology, spin-orbit torque, and dynamic symmetry breaking on the growth of magnetic stripe domains

J. A. Brock[1,*], D. Swinkels[1,2], B. Koopmans[2], and E. E. Fullerton[1]

[1] Center for Memory and Recording Research, University of California – San Diego, La Jolla, CA 92093 USA

[2] Department of Applied Physics, Eindhoven University of Technology, 5600 MB Eindhoven, The Netherlands

**Abstract**: We have performed an experimental and modeling-based study of the spin-orbit torque-induced growth of magnetic stripe domains in heavy metal/ferromagnet thin-film heterostructures that possess chiral Néel-type domain walls due to an interfacial Dzyaloshinskii-Moriya interaction. In agreement with previous reports, the stripe domains stabilized in these systems exhibit a significant transverse growth velocity relative to the applied current axis. This behavior has previously been attributed to the Magnus force-like skyrmion Hall effect of the stripe domain spin topology, which is analogous to that of a half-skyrmion. However, through analytic modeling of the in-plane torques generated by spin-orbit torque, we find that a dynamical reconfiguration of the domain wall magnetization profile is expected to occur - promoting motion with similar directionality and symmetry as the skyrmion Hall effect. These results further highlight the sensitivity of spin-orbit torque to the local orientation of the domain wall magnetization profile and its contribution to domain growth directionality.

## Introduction

In recent years, the dynamics of magnetic textures with chiral domain walls (DWs) have been the subject of intense research interest. In particular, it has been shown that chiral Néel-type DWs permit the efficient translation of magnetic domains in response to the spin-orbit torque (SOT) exerted by injected spin currents[1,2]. This scheme can easily be realized via the spin-Hall effect[3,4] when an electrical current passes through a heavy metal (HM) layer with large spin-orbit coupling adjacent to a ferromagnet (FM) layer.[5] Furthermore, the interfacial Dzyaloshinskii-Moriya interaction (iDMI) that develops at the FM/HM interface fosters a preference for chiral Néel-type DWs, with the strength and handedness of this interaction being determined by the chemical identity and quality of the interface.[6,7]

Besides promoting chiral Néel-type DWs, the iDMI also assists in stabilizing novel spin textures, such as magnetic skyrmions.[8,9,10,11,12] The quantized winding-like structure of skyrmions has been shown to give rise to several application-relevant properties, including a "topological protection" of the magnetic state and efficient SOT-induced motion.[13] At a more fundamental level, it has been theoretically and experimentally demonstrated that the non-trivial topological charge of skyrmions gives rise to a Magnus force-like dynamic action known as the skyrmion Hall effect, whereby a skyrmion driven into motion experiences a transverse deflection relative to the force exerted by the Slonczewski-like component of the SOT, $\vec{F}_{SL}$.[14,15,16,17] While the skyrmion Hall effect was first studied using skyrmions (with a topological charge $|Q| = 1$), there have been several recent reports of skyrmion Hall effect-like qualities to the SOT-induced growth of magnetic stripe domains.[18,19,20,21] By considering the tip of the stripe domain to be a half-skyrmion (with $|Q| = ½$), the deflection angle of a stripe domain relative to the force exerted by the SOT ($\theta$) has been understood in the framework of the skyrmion Hall effect using the expression:[18]

$$\theta \approx \tan^{-1}\left(\frac{2Q\lambda}{\alpha r}\right) \tag{1}$$

where $\lambda$ is the DW width, $\alpha$ is the Gilbert damping parameter, and $r$ is the half-skyrmion radius. While $r$ and $\lambda$ can be measured or calculated straightforwardly, measuring $\alpha$ can be more difficult, particularly in thin-film systems with strong perpendicular magnetic anisotropy (PMA). As such, $\alpha$ has typically been attributed an *a priori* value or used as a fitting parameter when interpreting the SOT-induced growth directionality of stripe domains.[19,20,20]

Recently, it has been shown that the dynamic reconfiguration of the DW magnetization profile that occurs during the magnetic field-driven expansion of stripe domains with chiral Néel-type DWs can give rise to novel growth asymmetries that cannot be predicted solely from static energy treatments.[22] A key component to understanding the growth behavior in such systems required accounting for the fact that a perpendicular magnetic field exerts an in-plane torque on the DW magnetization profile – twisting the Néel-type profile extant at static equilibrium to that of a chiral mixed Bloch-Néel-type DW.[23] By the same token, the in-plane torque associated with the Slonczewski-like effective magnetic field ($H_{SL}$) that arises when a spin current is injected into a ferromagnet[24,25] can lead to a similar reconfiguration of the DW magnetization profile.[26,27] Furthermore, given that both the magnitude and directionality of $H_{SL}$ are sensitive to the core magnetic orientation at each azimuthal position around a DW[28] (whereas field-driven growth asymmetries are sensitive to the relative energy of each position[29]), understanding SOT-induced growth asymmetries may be a more straightforward proposition.

Here, we report on a study of the SOT-induced growth of stripe domains in several HM/FM thin-film multilayers designed to possess an iDMI and a net SOT acting on the FM layers when a charge current is passed through the structure. Like previous reports, we experimentally find that the stripe domains exhibit a strong growth component transverse to the applied current axis. By accounting for the in-plane torques associated with SOT effective perpendicular fields, our modeling finds that the reconfigured DWs have magnetization profiles that favor growth with the same general directionality and symmetries observed experimentally.

**Experimental Techniques:**

Samples of the structure Ta (2)/ Pt (5)/ [Co (0.7)/ Ni (0.5)/ Pt (0.7)]$_3$/ Co (0.7)/ Ni (0.5)/ Ta (3) and Ta (2)/ Pt(5)/ [Co (0.5)/ Pt (0.7)]$_4$/ Co (0.5)/ Ta (3) (thicknesses in nm; henceforth referred to as the Pt/Co/Ni and Pt/Co samples, respectively) were grown using dc magnetron sputtering. All samples were grown on Si substrates with a 300 nm-thick thermal oxide coating using a 3 mTorr partial pressure of Ar and a sputtering power of 50 W. As Pt and Ta are known to possess oppositely signed spin-Hall angles,[30,31,32] the seeding and capping layers were chosen to promote a net SOT acting on the FM layers when a charge current is passed through the structure. The samples were fashioned into 50 µm or 20 µm-wide wires using conventional metal lift-off UV photolithography techniques; Ti/Au contacts were deposited at the wire ends by similar means. Domain growth was observed using a polar magneto-optic Kerr effect (MOKE) microscope, manufactured by Evico Magnetics. In the polar MOKE images, dark (light) contrast corresponds to out-of-plane (in-to-plane) oriented domains. The convention used to define the domain growth angle $\theta_{\text{growth}}$ relative to the conventional current direction is shown schematically in Figures 1a and 1b. Electrical current pulses were supplied to the samples using a MOSFET-gated power supply, while the current delivered to the samples was monitored using an oscilloscope. The room-temperature static magnetic properties of continuous films of each composition were determined using the vibrating sample magnetometry (VSM) technique, in both the out-of-plane (OOP) and in-plane (IP) geometries.

**Experimental Results**

The Pt/Co/Ni sample discussed in this report has previously been shown to exhibit a saturation magnetization $M_s$ and an effective perpendicular magnetic anisotropy (PMA) energy density $K_{\mathrm{eff}}$ of approximately 1000 kA/m and 500 kJ/m$^3$, respectively.[22] From Brillouin light scattering spectroscopy measurements reported elsewhere, the Pt/Co/Ni sample was also found to possess an iDMI energy density of $D \approx -0.6$ mJ/m$^2$ (with the negative sign of $D$ denoting a left-handed iDMI), which was sufficiently strong enough to stabilize Néel-type DWs (as evidenced by Lorentz transmission electron microscopy). Magnetization data for the Pt/Co sample (shown in Supplemental Figure 1) indicates $M_s$ and $K_{\mathrm{eff}}$ values of 890 kA/m and 415 kJ/m$^3$, respectively. Through measurements of the in-plane field-induced domain growth asymmetries present in a sample with fewer Pt/Co repetitions (where magnetic reversal occurs via large, circular domains), we estimate a $D \approx -0.42$ mJ/m$^2$ in our Pt/Co structures, as shown in Supplemental Note 1. While the Pt/Co/Pt motif would not be expected to exhibit an iDMI given its structural inversion symmetry, past works have shown that differences in the quality of the Pt/Co and Co/Pt interfaces can lead to behavior more associated with structural inversion asymmetry, like PMA and iDMI.[33,34] Both the Pt/Co/Ni and Pt/Co samples reverse their magnetization via dendritic stripe domains, as is typical for systems with multiple repetitions of the FM/HM motif.[35,36]

In Figure 1a, we show a polar MOKE image of the Pt/Co/Ni sample after several 10 ms-long current pulses with a current density of 3.5 x $10^{11}$ A/m$^2$ were applied along the $+\hat{x}$ direction indicated in Figure 1c. Before applying the current pulses, the sample magnetization was saturated into the sample plane ($-M_z$). Inspecting the domain growth relative to the points at which the domains are first nucleated (indicated by the blue dots), it can be seen that the growth component collinear to the long axis of the wire is parallel to the conventional current density $J$, as would be expected from domains with left-handed chiral Néel DWs subject to a positive net SOT.[32] Similar to previous studies of the SOT-induced growth of stripe domains, we observe that the domains are strongly pinned to the initial nucleation sites and that the width of SOT-nucleated/driven domains is noticeably smaller than the width of domains nucleated and driven solely using magnetic fields. These behaviors have previously been attributed to disparities between the driving and pinning forces acting on domains when DW motion is driven by magnetic fields versus SOT.[18,19,20]

Throughout Figure 1, it is also readily apparent that the reversal domains exhibit a significant growth component transverse to $J$. To define the angle at which the domains grow relative to $J$ ($\theta_{\mathrm{growth}}$), we use the angular conventions shown schematically within the insets of Figure 1. We note that while the same conventions are used for both $+M_z$ and $-M_z$ domains, $\theta_{\mathrm{growth}}$ is measured clockwise (counterclockwise) relative to the $+x$-axis ($-x$-axis) for $+J$ ($-J$). Using this convention, Figure 1a demonstrates that all of the magnetic domains within the track move at roughly the same growth angle $\theta_{\text{growth}}$ of 48.0° ± 1.6° relative to the *J*-axis. Given that the gyrotropic force that gives rise to the skyrmion Hall effect is expressed as $\overrightarrow{F_G} = -4\pi Q|v|(\hat{z} \times \vec{v})$ (where $v$ and $\hat{z}$ are the stripe domain expansion velocity and the unit vector pointing out of the film plane, respectively),[18] it can be seen that the transverse deflection observed in our samples does indeed match the directionality one would expect from the skyrmion Hall effect acting on the growth front of the stripe domain.

Reversing the polarity of the applied current (Figure 1b), it can be seen that the domain growth components, both collinear and transverse to the applied current axis, are reversed – as would be expected from chiral Néel-type DWs subject to a spin current of opposite polarization from that used in Figure 1a. If the sample is instead saturated in the $+M_z$ direction before applying electrical current pulses so that $-M_z$ domains are nucleated by the current pulses (Figures 1c and 1d) we find that the growth component collinear to $J$ is the same for $-M_z$ and $+M_z$ domains (which is the expected response for chiral Néel-type DWs), but the transverse growth component is in the opposite direction. Given that the sign of $Q$ is opposite for $+M_z$ and $-M_z$ domains, the sign change of the growth component transverse to $J$ is consistent with the skyrmion Hall effect. Furthermore, using the $\theta_{\text{growth}}$ conventions shown in Figure 1, we find that the absolute value of $\theta_{\text{growth}}$ does not vary significantly between these four permutations in the charge current and domain polarity. Similar growth trends with respect to domain polarity and the electrical current direction are observed in the Pt/Co samples (Supplemental Figure 3).

In Figure 2, we plot the dependence of $\theta_{\text{growth}}$ on current density $J$ and expansion velocity $|v|$ for $+M_z$ domains in the Pt/Co/Ni sample subjected to 10 ms-long current pulses. As previously noted in other works, we find that $\theta_{\text{growth}}$ is relatively constant across the $J$ (and hence, $|v|$) values attainable with our experimental setup.[19,20] While $\theta$ is expected to saturate in the limit of high $|v|$ – when the impact of pinning relative to the SOT driving force is insignificant[16], such behavior at lower speeds, where domain growth is strongly affected by thermally-activated pinning/depinning at defect sites[37,38] (*i.e.*, the creep dynamical regime) is unexpected. Given that the $J$ values accessible to us are only slightly above the threshold values needed to observe domain growth (and consequentially, $|v|$ is limited to under 200 μm/s), it is reasonable to believe that the domain growth reported here occurs within the creep dynamical regime. As such, the invariance of $\theta$ with respect to $J$ is not commensurate with the expectations of the skyrmion Hall effect in the creep regime.[16]

As shown in Figure 2b, we also find that both $|v|$ and $\theta$ increase when a dc magnetic field parallel to the domain polarity ($H_z$) is applied simultaneously with the electrical current pulses. The increase in $|v|$ with $H_z$ in the stripe domain morphology has previously been ascribed to an increase in the radially-divergent force that $H_z$ exerts on all azimuthal positions around the DW, which favors domain expansion.[20] In the same work, the increase in $\theta$ with $H_z$ was attributed to the fact that the net $\overrightarrow{F_{H_z}}$ (obtained via the integration of $\overrightarrow{F_{H_z}}$ around the perimeter of the DW) does not approach zero for the half-skyrmion structure. More specifically, the net $\overrightarrow{F_{H_z}}$ was shown to act in the $\vec{v}$-direction dictated by the balance of $\vec{F}_{SL}$ and $\vec{F}_G$. By this logic, the component of $\overrightarrow{F_{H_z}}$ parallel to $\overrightarrow{F_G}$ leads to an enhanced growth component transverse to $J$, yielding a larger $\theta$ for the same applied $J$ – an understanding that agrees with the trend demonstrated in Figure 2b.

**Analytical Modeling and Discussion**

Based on the earlier discussion of the forces that accompany SOT-based domain expansion, Equation 1 can be used to estimate the Gilbert damping $\alpha$ that would be required if the transverse domain growth exhibited by our samples was solely due to the skyrmion Hall effect. When evaluating Equation 1, the DW width λ is calculated from the relation $\lambda = \sqrt{A/K_{eff}}$, assuming that $A$ = 10 pJ/m (as is typical for FM/HM multilayers with these magnetic layer thicknesses). Using the static magnetic properties mentioned in the Experimental Results section, Equation 1 dictates that a Gilbert damping $\alpha$ of ~0.01 would be needed for the Pt/Co/Ni sample if the $\theta_{\text{growth}}$ of ~50° we observe arises entirely from the skyrmion Hall effect. While we have not performed an explicit measurement of $\alpha$ for our samples, past work has shown that an $\alpha$ value of 0.01 is upwards of one order of magnitude lower than what has been reported for FM/HM structures with similar compositions, degrees of perpendicular anisotropy, and thicknesses.[39,40] Furthermore, it should also be noted that Equation 1 was initially derived to describe stripe domain motion in an amorphous ferrimagnetic sample near its compensation temperature, where the domain expansion velocity approached 1 km/s, and thus, the pinning forces were insignificant relative to the driving force.[18] As such, the applicability of Equation 1 to stripe domains moving at the creep-scale velocities reported here and in other works is somewhat debatable. Based on these points, further exploration of the origins of the significant growth angles observed in our films is warranted.

When interpreting the growth behavior of stripe domains driven by SOT, the DW magnetization profile is typically assumed to be that which exists at static equilibrium. This static configuration can be determined by minimizing the expression for the DW energy density $\sigma$, given as:[41]

$$\sigma(\theta,\varphi) = 4\sqrt{AK_{eff}} + \frac{\ln 2}{\pi} t_f \mu_0 M_s^2 \cos^2(\varphi - \beta) - \pi|D|\cos(\beta) \qquad (2)$$

where $\mu_0$ is the vacuum permeability, $t_f$ is the total ferromagnetic thickness, and $A$ is the exchange stiffness parameter. We define the DW magnetic orientation ($\varphi$) and azimuthal position ($\beta$) angles counterclockwise relative to the $+\hat{x}$-direction, as demonstrated in Figure 3a. For the Pt/Co/Ni sample, it is known that the balance between the terms in this expression predicts the stabilization of left-handed Néel-type DWs in the absence of any external stimulus,[22] as illustrated by the half-skyrmion structure shown in Figure 3b.

As detailed in previous works[18,20,42,43,44], four forces are predicted to govern the SOT-induced dynamics of Néel-type DWs: The aforementioned net $\overrightarrow{F_{SL}}$ and $\overrightarrow{F_G}$, as well as a viscous drag force ($\overrightarrow{F_D}$) and the tension arising from pinning effects ($\overrightarrow{F_T}$). When DW motion occurs within the flow regime, the growth directionality in response to SOT is primarily determined by the steady-state balance of $\overrightarrow{F_G}$ relative to $\overrightarrow{F_{SL}}$, with $\overrightarrow{F_D}$ and $\overrightarrow{F_T}$ acting to limit the magnitude of the velocity. In the creep dynamical regime, however, it is believed that $\overrightarrow{F_G}$ should be insignificant relative to $\overrightarrow{F_{SL}}$.[16,37,38] Additionally, we note that the $\overrightarrow{H_{SL}}$ at each azimuthal position (which in turn gives rise to $\overrightarrow{F_{SL}}$) can be determined using the expression:

$$\mu_0 \vec{H}_{SL} = \frac{\mu_0 \hbar \theta_{SH}}{2|e| M_s t_f} |J| (\widehat{m}_\theta \times \vec{\sigma}) \qquad (3)$$

where $\hbar$ is the reduced Planck constant, $\theta_{SH}$ is the spin-Hall angle, $e$ is the charge of the electron, and $\vec{\sigma} = \hat{z} \times \hat{\jmath}_e$ is the spin-polarization axis (where $\hat{\jmath}_e$ is the unit vector in the electron flow direction). Assuming $\theta_{SH}$ is positive (which is reasonable considering the Pt/FM/Ta structure of our samples[30,31,32]), a half-skyrmion with chiral Néel-type DWs driven within the creep regime would have the force interrelationship and expansion velocity direction shown schematically in Figure 3b. Critically, given that the net $\overrightarrow{F_{SL}}$ acting on a half-skyrmion is determined by summing $H_{SL}$ azimuthally along the DW and the magnitude and direction of $H_{SL}$ is determined by the orientation of $\varphi$ relative to $J$, a half-skyrmion with horizontal mirror symmetry (like that shown in Figure 3b) should experience a net $\overrightarrow{F_{SL}}$ that is collinear to $\hat{x}$.

Recently, it has been demonstrated that when stripe domains with chiral Néel-type DWs are driven by a perpendicular magnetic field, the reconfiguration of the DW magnetization profile induced by the associated in-plane torques can dramatically impact the directionality of domain growth in the presence of an energy symmetry-breaking in-plane magnetic field.[22] Since $H_{SL}$ behaves as a perpendicular magnetic field that acts on the DW core magnetization, it should be capable of driving similar reconfigurations of the DW magnetization profile.[23] The steady-state dynamic reconfiguration of the DW magnetization profile (relative to that extant at static equilibrium) is dictated by the condition:[23]

$$\dot{\varphi} = \frac{|\gamma|}{1+\alpha^2}\left[-\alpha\left(\frac{\mu_0\sigma_\varphi}{2\lambda\mu_0 M_s} - \frac{\pi}{2}\mu_0 H_{FL}\right) + \frac{\pi}{2}\mu_0 H_{SL}\right] = 0 \tag{4}$$

where $|\gamma|$ is the gyromagnetic ratio, subscripts of $\sigma$ indicate a partial derivative of Equation 2, and $H_{FL}$ represents the field-like component of the SOT (which acts as an effective magnetic field collinear to the $+\hat{y}$ direction). For a film with PMA, $H_{FL}$ can modify the DW magnetization profile, but would not be expected to drive domain expansion. Furthermore, it is believed that the orientation of $H_{FL}$ relative to $J$ is determined by the composition of the HM layers that generate the SOT, not the DW configuration of the FM layer.[1] For the Pt seeding and Ta capping layer used in this work, it has been shown that the magnitude of $H_{FL}$ can be of a similar magnitude as $H_{SL}$.[32]

A schematic depiction of the steady-state dynamic reconfiguration of the DW magnetization profile predicted by Equation 4 is shown in Figure 3c. Comparing Figures 3b and 3c, it can be seen that the torque exerted by $H_{SL}$ induces a twisting of the half-skyrmion's DW magnetization profile relative to that which exists at static equilibrium – effectively breaking the horizontal mirror symmetry of the Néel chirality. For the domain structure shown in Figure 3c, this torque promotes a DW profile that contains a mixture of left-handed Néel and right-handed Bloch components. In line with the previous discussion, the degree of twisting at each azimuthal position is proportional to the scalar product of the magnetic orientation at static equilibrium (*i.e.*, the $\beta(\varphi)$ profile shown in Figure 3b) and the electron flow direction. Most notably, the azimuthal position where $|H_{SL}|$ is strongest (*i.e.*, where $\varphi$ = 180°) is no longer at $\beta$ = 0° but has instead been shifted counterclockwise about the perimeter of the half-skyrmion. Because of this reconfiguration, the net $\overrightarrow{F_{SL}}$ will no longer be purely collinear to $\hat{x}$ but will also possess a component in the $\hat{y}$-direction as well – as has been demonstrated for SOT-induced domain growth when a magnetic field is applied along $\hat{y}$ to realize a reconfiguration of the DW magnetization profile.[20]

To compare this theoretical understanding of how in-plane torques impact the SOT-induced growth behavior of stripe domains with our experimental results, we have used Equation 4 to analytically determine the dynamic steady-state magnetization profile of the Pt/Co/Ni sample induced by spin-orbit torque (determined using Equation 3). When performing these calculations we employ the static magnetic properties stated in the Experimental Results section along with a Gilbert damping of 0.2,[22] a $\theta_{SH}$ value of 0.2,[32] and with the assumption that $|H_{FL}| = |H_{SL}|$ and is oriented along the $+\hat{y}$ ($-\hat{y}$) direction when $J$ is applied in the $+\hat{x}$ ($-\hat{x}$) direction.[1,32] Using these parameters, the steady-state DW core magnetic orientation at each azimuthal position around the half-skyrmion [$\varphi(\beta)$] was obtained for a variety of $J$-values, as shown in Figure 4a. Referencing the conventions for $\varphi$ and $\beta$ shown in Figure 3a, it can be seen that the SOT effective fields successively twist the magnetization away from the static equilibrium configuration in a clockwise manner. Based on this SOT-induced reconfiguration of $\varphi(\beta)$, we have also calculated the $H_{SL}$ experienced at each azimuthal position on the half-skyrmion's perimeter using Equation 3, as shown in Figure 4b. To aid in comparing between curves calculated using different $J$ values, we have normalized $H_{SL}$ to the maximum value present in each respective curve. As was observed in the $\varphi(\beta)$ curves, we find that $H_{SL}(\beta)$ is no longer symmetric about $\beta = 0°$ when $J > 0$.

To understand how these asymmetries of the $\varphi(\beta)$ and $H_{SL}(\beta)$ profiles impact the growth directionality of stripe domains, we have employed the Thiele equation-based technique proposed in Ref. 20. Using our calculated $H_{SL}(\beta)$ profiles to obtain the $x$ and $y$ components of the net $F_{SL}$ via azimuthal integration, a directionality can be assigned to the net $F_{SL}$ acting on the tip of a stripe domain. As previously discussed, we assume that our experiments occur in the creep dynamical regime; as such, we do not consider the impacts of the skyrmion Hall effect when determining the favored growth direction. In Figure 4c, we show the analytically predicted $\theta_{growth}$ of half-skyrmions for different permutations in domain polarity and magnitude/polarity of $J$. While the analytical modeling suggests that $\theta_{growth}$ should show a strong dependence on $|J|$, our experimentally determined $\theta_{growth}$ values (also shown in Figure 4c) are rather invariant with $|J|$.

In previous studies that attributed the directionality of SOT-induced half-skyrmion growth to the skyrmion Hall effect, a similarly quizzical invariance of $\theta_{\mathrm{growth}}$ with $|J|$ has been noted.[19,20] Even when accounting for a reconfiguration of the DW magnetization profile to describe the SOT-induced growth directionality of bubble-type domains confined to nanoscale geometries, $\theta_{\mathrm{growth}}$ was found to be relatively static even when the growth velocity was varied over three orders of magnitude.[45,46] While our analytical approach correctly predicts the symmetries and general directionality of SOT-induced half-skyrmion growth while assuming realistic Gilbert damping values, the reconfiguration mechanism of the DW magnetization profile applied here stems from the dynamics of driven DWs, and our analytical modeling does not account for the fact that DW motion within the creep regime proceeds via spurts of motion separated by pinning at defect sites on the order of several microseconds. As such, while our approach correctly predicts the symmetries and general directionality of SOT-induced half-skyrmion growth, further work is needed to understand the balance of forces and the role that pinning plays in governing the observed independence between $|\theta_{\mathrm{growth}}|$ and $|J|$ in the creep dynamical regime.

**Conclusion**

We have performed an experimental and analytic study of the growth directionality of dendritic stripe domains in response to spin-orbit torque in several ferromagnet/heavy metal multilayer systems. In line with previous reports[18,19,20], it was found that the stripe domains exhibit a significant transverse growth component relative to the applied current axis and that this growth component exhibits specific symmetries with respect to the domain polarity and electric current direction. While previous treatments considered the tips of stripe domains to be half-skyrmions, thus attributing these growth behaviors to the skyrmion Hall effect, we have analytically shown that the in-plane torques associated with the spin-orbit torque effective field can promote growth directionality with the same symmetries as the skyrmion Hall effect, well within the creep dynamical regime. Overall, our findings illustrate that the transverse growth of magnetic domains in response to spin-orbit torque can arise from spin topology-dependent behaviors other than the skyrmion Hall effect.

**Acknowledgments**

JAB and EEF acknowledge support from the National Science Foundation, Division of Materials Research (Award #: 2105400). This work was performed in part at the San Diego Nanotechnology Infrastructure (SDNI) of the University of California – San Diego, a member of the National Nanotechnology Coordinated Infrastructure, which is supported by the National Science Foundation (Grant #: ECCS-1542148).

[*]Present address: Laboratory for Mesoscopic Systems, Department of Materials, ETH Zurich, 8093 Zurich, Switzerland and Laboratory for Multiscale Materials Experiments, Paul Scherrer Institute, 5232 Villigen PSI, Switzerland

[44] A.A. Thiele, *Phys. Rev. Lett.* **30**, 230 (1972).

[45] C.K. Safeer, E. Jue, A. Lopez, L. Buda-Prejbeanu, S. Auffret, S. Pizzini, O. Boulle, I.M. Miron, and G. Gaudin, *Nature Nanotech.* **11**, 143 (2016).

[46] E. Martinez, O. Alejoa, M.A. Hernandez, V. Raposo, L. Sanchez-Tejerina, and S. Moretti, *Appl. Phys. Expr.* **9**, 063008 (2016).

**Figures:**

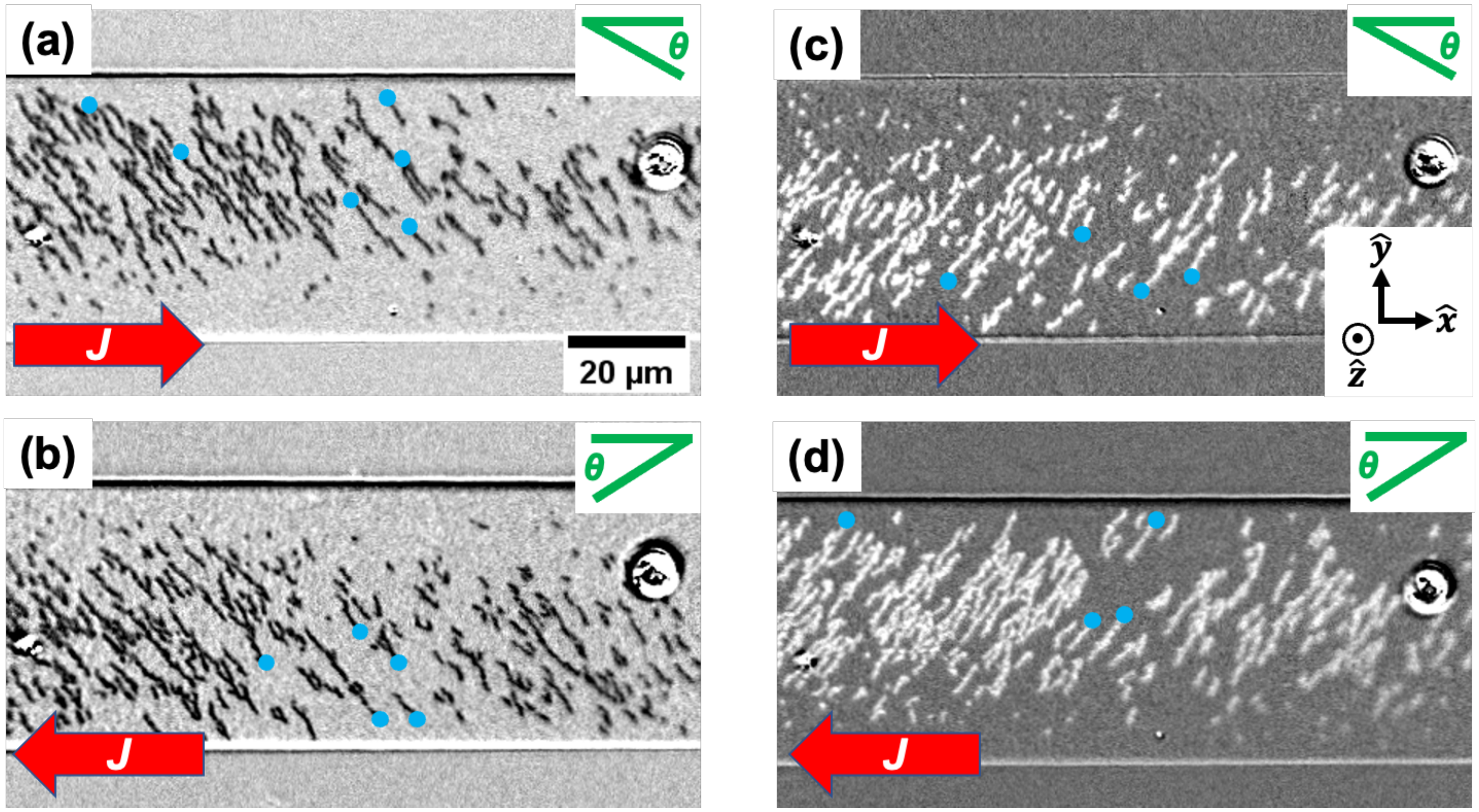

**Figure 1**: Polar MOKE images of the Co/Ni/Pt sample after several 10 ms-long current pulses of current density 3.5 x $10^{11}$ A/$m^2$ were applied in zero magnetic field. The conventional current flow direction is indicated in the lower-left corner of each subfigure, whereas the conventions used to define $\theta_{growth}$ for different permutations in the current and domain polarity are shown in the upper-right corners. The lower-right inset of (c) depicts the coordinate axes used throughout this work. Select locations at which reverse domains are initially nucleated when applying electrical current pulses are indicated by blue dots.

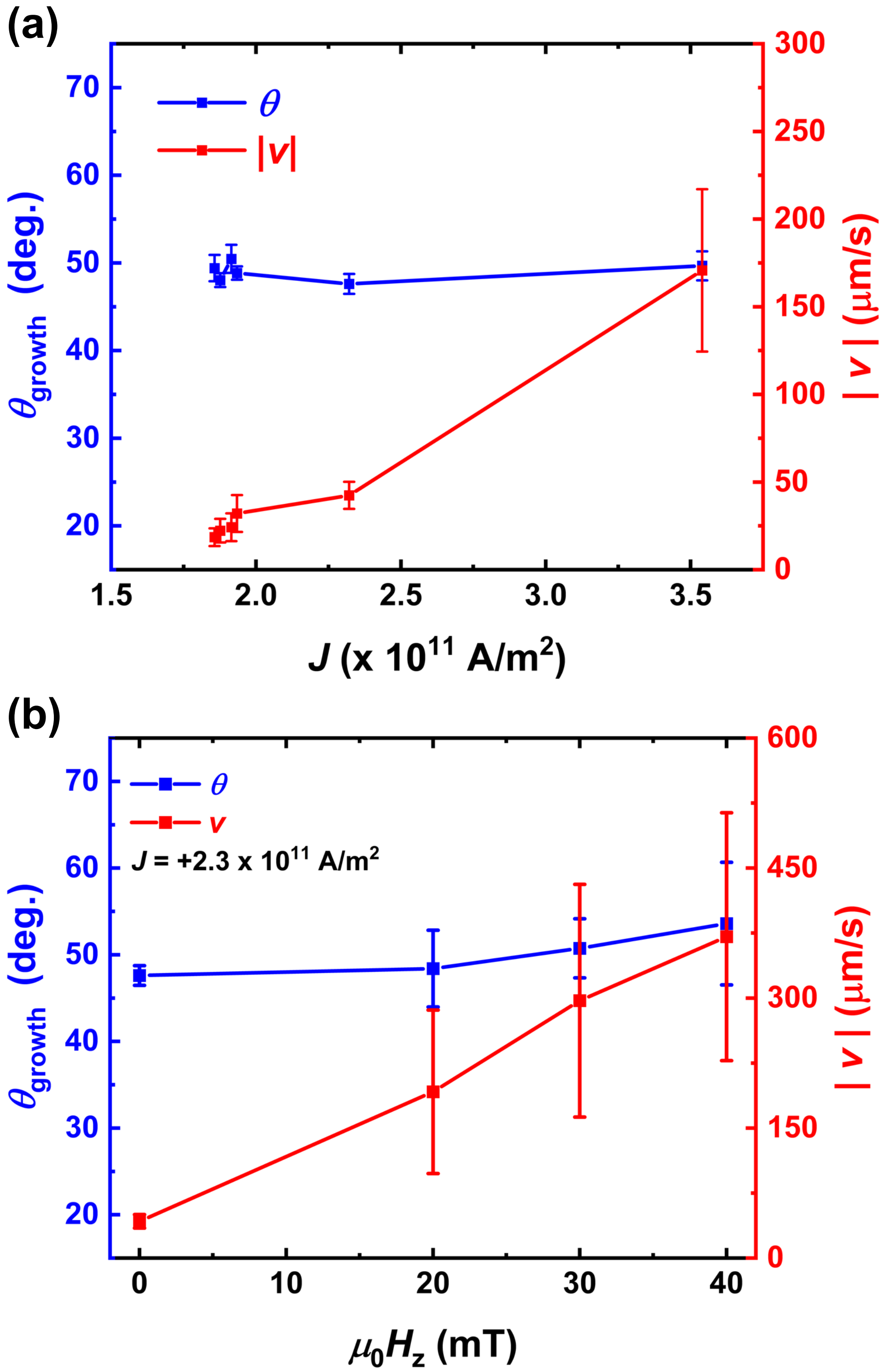

**Figure 2:** For the Co/Ni/Pt sample, (a) the growth direction $\theta_{growth}$ and the magnitude of the expansion velocity $|v|$ for the Co/Ni/Pt sample as a function of applied current density $J$. (b) $\theta_{growth}$ and $|v|$ as a function of perpendicular bias field $H_z$ for a fixed current density of 2.3 x $10^{11}$ A/m$^2$. The data shown here corresponds to the domain and current polarity demonstrated in Figure 1a. The $\theta_{growth}$ values shown were determined by averaging the growth directionality of 25 different reversal domains, whereas the error bars represent the standard error.

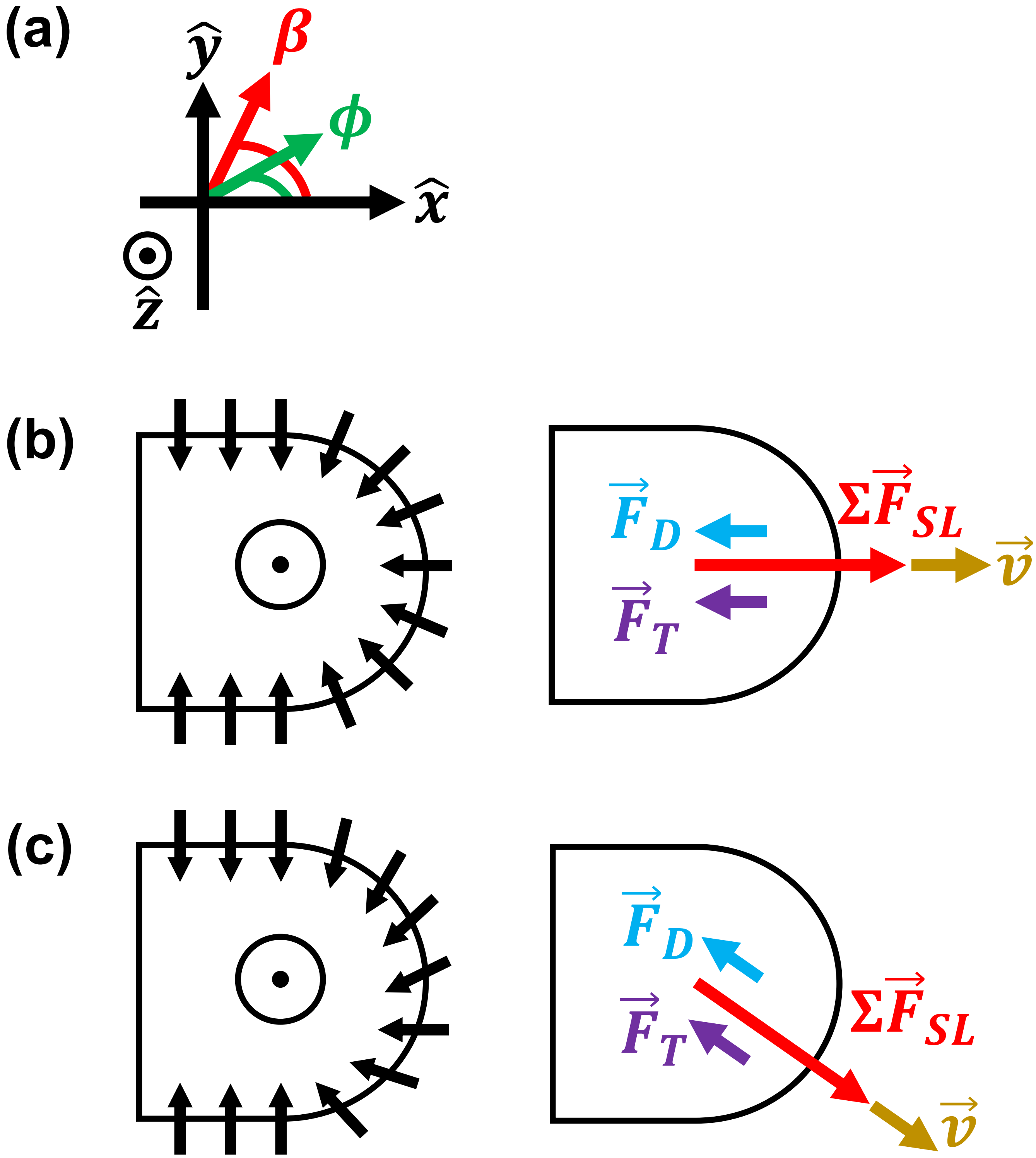

**Figure 3:** (a) The conventions used throughout this work to define the azimuthal angle around the perimeter of a half-skyrmion ($\beta$) and the DW core magnetization orientation ($\varphi$). (b,c) Schematic depiction of the DW magnetization profile and SOT force interrelationship in a half-skyrmion assuming the DW magnetization profile that exists at static equilibrium (b) and that resulting from a dynamic reconfiguration of the DW magnetization profile (c).

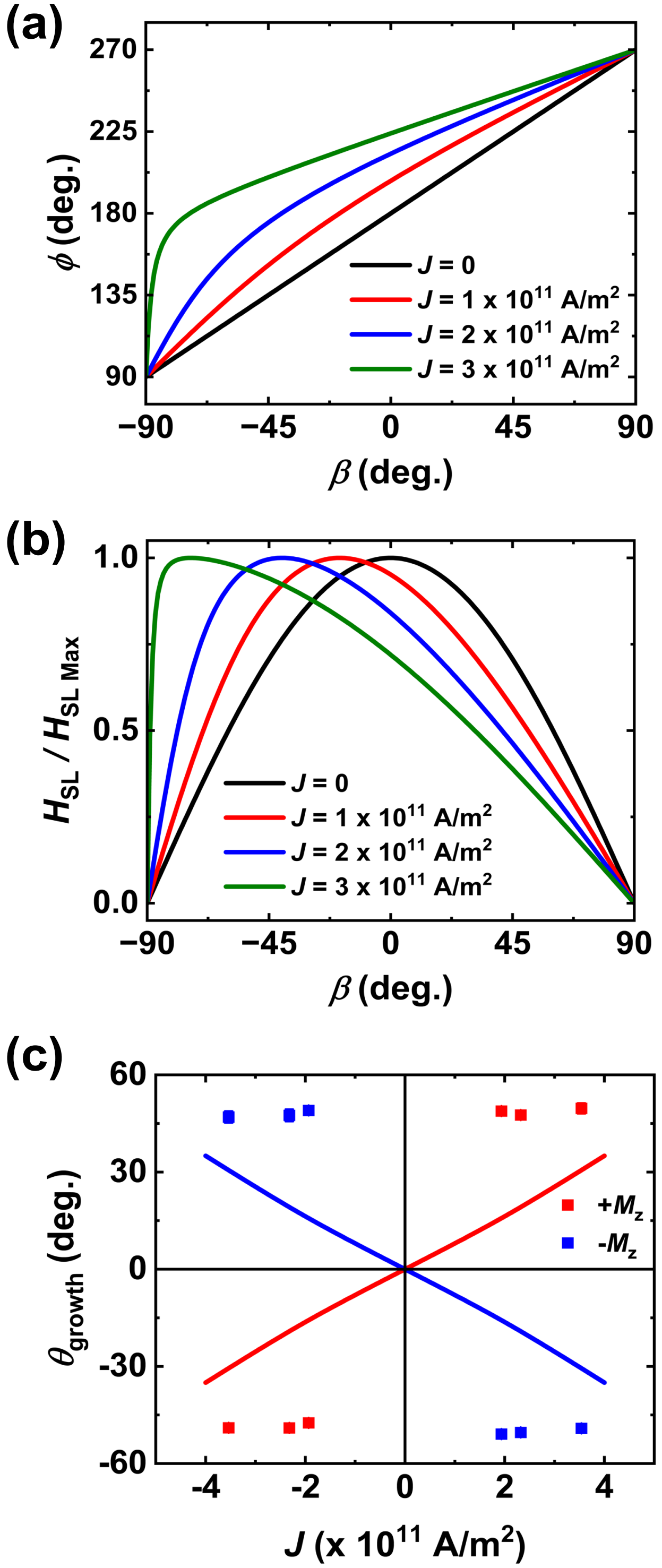

**Figure 4:** For the Co/Ni/Pt sample: (a) Analytically calculated DW core magnetization orientation $\varphi$ and (b) normalized Slonczewski effective perpendicular field $H_{SL}$ as a function of azimuthal position $\beta$ on a $+M_z$ half-skyrmion for several current densities. In (a) and (b), the conventional current direction is along the $+x$ axis indicated in Figure 1c. (c) Experimentally (square markers) and analytically (lines) determined favored growth direction $\theta_{growth}$ of half-skyrmions for different permutations in the current density magnitude/directionality and domain polarity. $\theta_{growth}$ is defined using the conventions shown in the insets of Figure 1.

**Supplemental Information:**

**Supplemental Figures:**

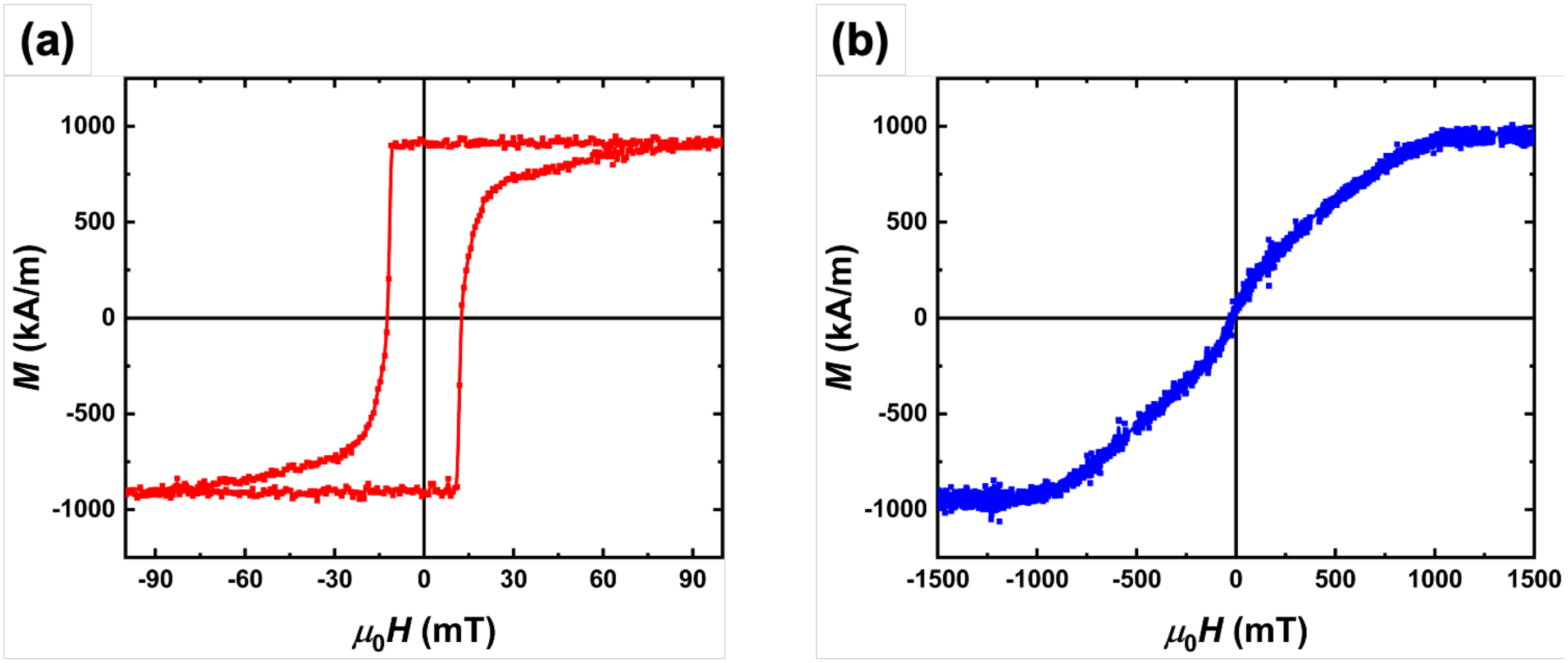

**Figure S1:** Room temperature hysteresis loops for the Pt/Co sample, collected in the (a) out-of-plane and (b) in-plane geometry using the vibrating sample magnetometry (VSM) technique.

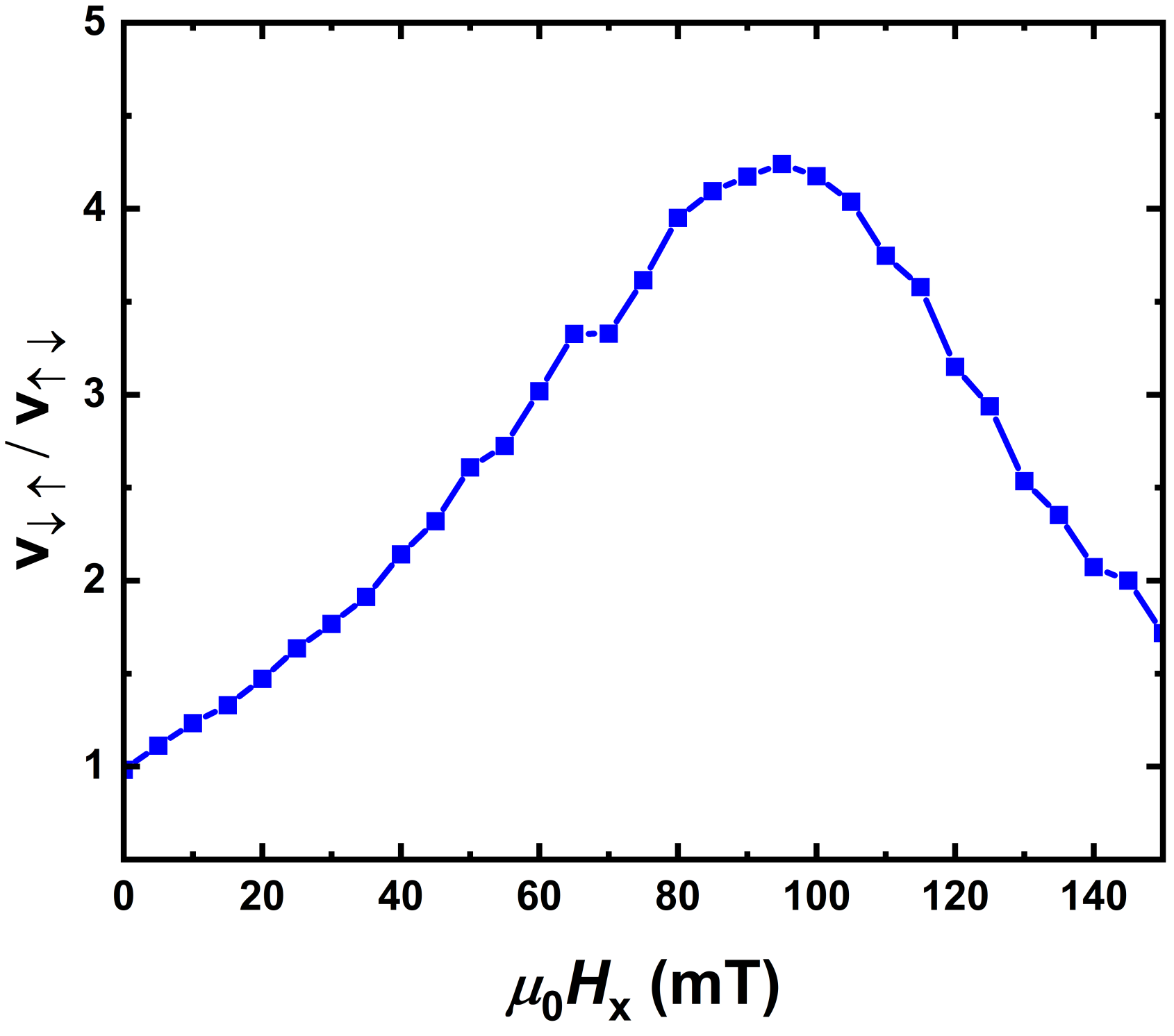

**Figure S2:** Ratio of the expansion velocity $v$ of a ↓↑ domain wall over that of a ↑↓ domain wall in a Co/Pt sample as a function of in-plane magnetic field $\mu_0 H_x$. the data shown was calculated from polar MOKE images collected after applying 5 ms-long, 15 mT-strong perpendicular magnetic field pulses. Identifying the peak in $v_{\downarrow\uparrow}$ / $v_{\uparrow\downarrow}$ as the iDMI effective field, an iDMI energy density of approximately -0.4 mJ/m$^2$ was calculated using techniques described in Ref. 28 of the main text.

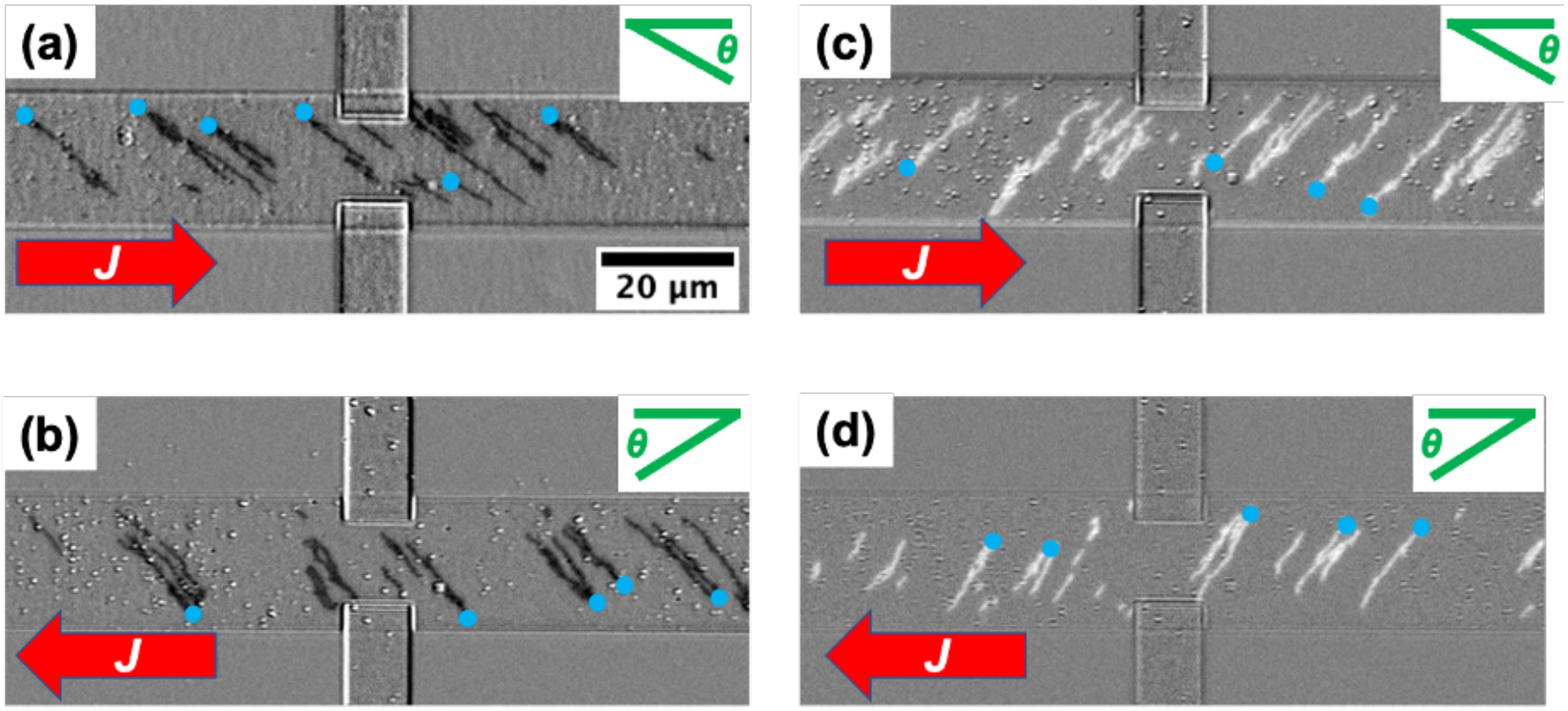

**Figure S3:** Polar MOKE images of the Co/Pt sample after several 10 ms-long current pulses of current density 5.2 x $10^{11}$ A/$m^2$ were applied in zero magnetic field. The conventional current flow direction is indicated in the lower left corner of each subfigure, whereas the convention used to define $\theta_{growth}$ is shown in the upper right corners.